\documentclass[a4paper,11pt]{article}
\usepackage[dvipsnames]{xcolor}
\usepackage{jheppub} 
\usepackage[T1]{fontenc} 
\usepackage{parskip}
\usepackage{cleveref}

\title{Effective Gravitational Couplings of Higher-Rank Supersymmetric
Gauge Theories}

\author[a]{Renjan Rajan John,}
\author[b,c]{Sujoy Mahato,}
\author[d]{and Madhusudhan Raman}

\affiliation[a]{School of Pure and Applied Physics\\
  Mahatma Gandhi University, Kottayam, Kerala 686 560, India}
\emailAdd{renjanjohn@mgu.ac.in}
\affiliation[b]{The Institute of Mathematical Sciences\\
IV Cross Road, C. I. T. Campus, Taramani, Chennai 600 113, India}
\affiliation[c]{Homi Bhabha National Institute\\
Training School Complex, Anushakti Nagar, Mumbai 400 094, India}
\emailAdd{sujoymahato@imsc.res.in}
\affiliation[d]{Instituto de F\'isica Te\'orica, Universidade Estadual
Paulista\\ R. Dr. Bento Teobaldo Ferraz 271, S\~ao Paulo 01140-070, Brazil\\}
\emailAdd{madhusudhan.raman@unesp.br}

\abstract{When placed on four-manifolds, $ \mathcal{N} = 2 $ gauge
  theories couple to topological invariants of the background via two
  functions $ A $ and $ B $. General considerations allow for these
  functions to be fixed in terms of the Coulomb moduli and other
  parameters in the theory, but only up to multiplicative factors
  about which little is known. We extend earlier work on the
  microscopic study of these functions in the $ \Omega $-background to
  $ \mathcal{N} = 2 ^{\star } $ gauge theories with higher-rank
  $ \mathrm{U}(N) $ gauge groups. We complement this analysis by
  carrying out a perturbative study of these functions. This allows us
  to determine the manner in which these multiplicative factors scale
  with the rank of the gauge group and the mass of the adjoint
  hypermultiplet.}

\begin{document}
	
\maketitle

\raggedbottom

\section{Introduction}
\thispagestyle{empty}

\subsection*{Gauge Theories on Four-Manifolds}
The quest to understand the strong-coupling dynamics of
four-dimensional gauge theories has been a steady source of motivation
for many theoretical physicists over the last
half-century. Significant among the numerous developments in this area
of research is the seminal work of Seiberg and Witten
\cite{Seiberg:1994rs,Seiberg:1994aj}, where holomorphy and duality
were used to essentially solve a strongly coupled (supersymmetric)
gauge theory.

Let us briefly recall the form that this solution takes. At low
energies, the Coulomb moduli space of these theories is parametrised
by a set of gauge-invariant order parameters
$ \left\lbrace u _{i} \right\rbrace $ and its dimension is the rank
$ r $ of the gauge group.\footnote{Throughout this paper, any sums or
  products over lowercase Latin indices will be understood to run over
  the set $ \{1,\cdots , r\} $.} A generic point on the Coulomb moduli
space finds the gauge group broken down to its maximal torus, and the
low-energy effective theory is one of $ r $ Abelian vector multiplets
controlled by a holomorphic function of the $ \mathcal{N} = 2 $
superfield called the prepotential. This prepotential receives
$ 1 $-loop exact perturbative corrections and an infinite series of
corrections due to instantons, whose contributions may be computed
using an algebraic curve. In essence, the Seiberg-Witten solution
established an equivalence between the quantum moduli space of a
supersymmetric gauge theory and the moduli space of an algebraic
curve, which in turn paved the way for a complete solution specifying
the low-energy dynamics of the theory.

Much subsequent attention has been paid to $ \mathcal{N} = 2 $
theories living at fixed points of renormalisation group flows,
i.e.~superconformal field theories with eight supercharges, henceforth
abbreviated $ \mathcal{N} = 2 $ SCFTs. Of particular interest in the
study of these theories is the determination of their conformal
central charges $ \texttt{a} $ and $ \texttt{c} $. Cognisant of the
limitations of earlier (more indirect) methods that leveraged
S-duality or holography to compute these quantities, Shapere and
Tachikawa \cite{Shapere:2008zf} proposed a more direct and general
method: since the conformal central charges characterise the response
of an $ \mathcal{N} = 2 $ SCFT to a metric perturbation, they proposed
to put these theories on curved four-manifolds using a topological
twist and take advantage of the relationship between the
$ \mathrm{U}(1)_{R} $ anomalies of $ \mathcal{N} = 2 $ gauge theories
and their topologically twisted counterparts.

Once again, it will be useful to --- if only briefly --- recall some
details. (We refer the reader to \cite{Shapere:2008zf} for a more
comprehensive discussion.) On one hand, we have the topological
twisting of an $ \mathcal{N} = 2 $ theory \cite{Witten:1988ze}, which
in the language of supergravity corresponds to switching on an
$ \mathrm{SU}(2)_{R} $ gauge field and setting it equal to the
self-dual part of the spin connection. This has the happy consequence
of leaving one component of the supercharge to transform as a scalar,
whose cohomology may subsequently be used to define physical
operators. On the other hand, the theories we consider have an
$ \mathcal{N}=2 $ superconformal algebra, whose $ R $-current anomaly
equation is specified in terms of the conformal central charges
$ \texttt{a} $ and $ \texttt{c} $. These two facts put together
\cite{Anselmi:1997am,Anselmi:1997ys,Kuzenko:1999pi} can be used to
relate the integrated $ \mathrm{U}(1)_{R} $ anomaly $ \Delta R $ of the vacuum
to a linear combination of topological invariants of the four-manifold
with coefficients that are functions of the conformal central charges:
\begin{equation}
  \label{eq:R-anomaly-1}
  \Delta R = 2 (2 \texttt{a} - \texttt{c}) \, \chi + 3c \, \sigma \ .
\end{equation}
In terms of the curvature $ 2 $-form $ \mathrm{R} $ and its dual
$ \tilde{\mathrm{R}} $, the topological invariants $ \chi $ and
$ \sigma $ above are the Euler characteristic
\begin{equation}
  \label{eq:euler-characteristic-density}
  \chi = \frac{1}{32 \pi ^{2}} \int \operatorname{tr} \mathrm{R} \wedge \tilde{\mathrm{R}} \ , 
\end{equation}
and the signature
\begin{equation}
  \label{eq:signature-density}
\sigma = \frac{1}{24 \pi ^{2}} \int \operatorname{tr} \mathrm{R} \wedge \mathrm{R} \ .
\end{equation}

The path integral of topologically twisted $ \mathcal{N} = 2 $ gauge
theories on four-manifolds appears, too, in the physical approach to
Donaldson invariants \cite{Witten:1988ze,Moore:1997pc}. Here, the
generating function of Donaldson invariants receives contributions
from the so-called $ u $-plane integral, an integral over the Coulomb
branch.\footnote{There is also a contribution from points in the
  Coulomb moduli space associated to extra massless particles, which
  we will not discuss here.} Let $ X $ be a four-manifold and
$ b _{p} $ be the dimension of the space of harmonic $ p $-forms on
$ X $. When referring to forms/spaces associated to (anti-)self-dual
forms we will use the superscripts $ (\pm) $. Now, for the case of
$ b _{1}(X) = 0 $ and $ b _{2} ^{+} (X) = 1 $, the $ u $-plane
integral takes the form
\begin{equation}
  \label{eq:u-plane-integral}
Z _{u} \sim  \int [\mathrm{d} a \, \mathrm{d} \bar{a}] \, \underline{A(u)^{\chi } B(u)^{\sigma }} \, \Psi \ .
\end{equation}
up to some normalisation that renders the integral dimensionless. We
have written the $ u $-plane integral with reference to a specific
choice of electric special coordinates. The (underlined) measure factor tells
us how the low-energy effective theory couples to the topological
invariants of the four-manifold we saw in
\cref{eq:euler-characteristic-density,eq:signature-density}. In terms
of the dimensions of spaces of harmonic forms, we recall that these
are simply $ \chi = \sum_{i} (-1)^{i} b _{i} $ and
$ \sigma = b _{2} ^{+} - b _{2} ^{-} $. Finally, the term $ \Psi $ is
associated to the photon partition function of the low-energy
effective theory and takes the form of a lattice theta function and
can depend on insertions.

The $ u $-plane integral is significant not just to the study of
four-manifold invariants, but also to $ \mathcal{N}=2 $ SCFTs, as one
of the key results established by \cite{Shapere:2008zf} was the direct
relationship between the central charges $ \texttt{a} $ and
$ \texttt{c} $ on one hand, and the functions $ A $ and $ B $ on
the other. The argument here relies on $ R $-symmetry restoration at
superconformal fixed points and goes as follows: say that the
functions $ A $ and $ B $ are (somehow) already determined, and
imagine sitting at a point on the Coulomb branch near a superconformal
fixed point. The low-energy effective theory is one of $ r $ vector
multiplets and (possibly, say) $ h $ neutral hypermultiplets, which
contribute $ (\chi + \sigma )/2 $ and $ \sigma / 4 $ respectively to
the integrated $ R $-charge. If we now imagine moving to the nearby
superconformal fixed point, the contribution to the
$ \mathrm{U}(1)_{R} $ anomaly of fields that go massless as we do so
must be added in --- this is precisely the $ R $-charge of the measure
factor in \cref{eq:u-plane-integral}. The total integrated
$ \mathrm{U}(1) _{R} $ anomaly is now
\begin{equation}
  \label{eq:R-anomaly-2}
\Delta R = \chi \, R(A) + \sigma \, R(B) + r \left( \frac{\chi + \sigma }{2}  \right) + h \, \frac{\sigma }{4} \ .
\end{equation}
Together, \cref{eq:R-anomaly-1,eq:R-anomaly-2} determine
$ \texttt{a} $ and $ \texttt{c} $ for any $ \mathcal{N} = 2 $ SCFT
that lives on the Coulomb branch of some $ \mathcal{N} = 2 $ gauge
theory in four dimensions. Given the central importance of the
functions $ A $ and $ B $ to the above argument, our attention
in this paper will be focused on them.

\subsection*{The Functions $ A $ and $ B $}

The requirement of modular invariance of the path integral coupled
with the observation that the $ u $-plane integral measure has an
anomaly under change of electric coordinates that must be compensated
for by the measure factor is in fact sufficient to constrain the form
of the $ A $ function. Noting further that all free massless states
contribute to $ \texttt{c} $ allows us to conclude that $ B $ should
vanish on the loci corresponding to them. For a generic
$ \mathcal{N} = 2 $ gauge theory, then, the functions $ A $ and $ B $
are expected
\cite{Witten:1995gf,Moore:1997pc,Losev:1997tp,Marino:1998bm,Marino:1998tb}
to take the form:
\begin{equation}
  \label{eq:A-and-B-general-forms}
A = \alpha \left( \operatorname{det} \frac{\mathrm{d} u _{i}}{\mathrm{d} a _{j}}  \right)^{1/2} \quad \mathrm{and} \quad B = \beta \Delta ^{1/8} \ .
\end{equation}
In the above expression, $ \Delta $ is the ``physical discriminant''
which may in general be different from the usual ``mathematical
discriminant'' of the Seiberg-Witten curve, for reasons clarified in
\cite{Shapere:2008zf}. First, the form of the Seiberg-Witten curve is
not unique, and its different forms (with different mathematical
discriminants as normally defined) may still yield the same low-energy
physics. Second, the $ B $ function is sensitive to massless
states, which do not necessarily correspond to cycles of a
Seiberg-Witten curve.

On general grounds, \cref{eq:A-and-B-general-forms} is as good as one
can do, and the constants $ \alpha $ and $ \beta $ cannot be fixed
from such general considerations. They may in general depend on the
masses of hypermultiplets, or the cut-off scale $ \Lambda $. While
there are conjectures for the $ N $-dependence of these coefficients
for $ \mathrm{SU}(N) $ super-Yang-Mills due to \cite{Marino:1998bm},
and arguments that $ \alpha $ and $ \beta $ are mass-independent for
the case of asymptotically free theories due to \cite{Marino:1998tb},
almost nothing else is known about the coefficients $ \alpha $ and
$ \beta $, except in the case of $ \mathrm{SU}(2) $ super-Yang-Mills
theory, where one can compare with Donaldson invariants.

In particular, to determine $ \alpha $ and $ \beta $ one needs to
specialise to a particular theory and a specific choice of
gravatational background. This specialisation will also provide us
with an opportunity to locate our motivations in recent and
interesting work on these questions.

\subsection*{Gravitational Background}
Our work is a continuation and extension of the work of
\cite{Manschot:2019pog} in two respects. Like them, our choice of
gravitational background will be the $ \Omega $-background of
$ \mathbb{C}^{2} $ (equivalently, $ \mathbb{R}^{4} $), familiar from
the equivariant localisation computations of
\cite{Nekrasov:2002qd,Nekrasov:2003rj}. Here, the scalar supercharge
constructed via the topological twisting procedure we described
earlier is not nilpotent but squares instead to an isometry of the
$ \Omega $-background, which in this case is independent rotations in
the two $ 2 $-planes that are parametrised by the complex numbers
$ (\epsilon _{1}, \epsilon _{2}) $. The path integral of an
$ \mathcal{N} = 2 $ gauge theory on such a background reduces to the
computation of equivariant integrals on the moduli space of
instantons. In the flat-space limit, one recovers using this
technology the low-energy effective prepotential, allowing for a
first-principles verification of the Seiberg-Witten solution.

The $ \Omega $-deformed partition function contains much more
information than just the prepotential; since the Euler characteristic
and signature of this background are
\begin{equation}
\chi \left(\mathbb{C} ^{2}\right) = \epsilon _{1} \epsilon _{2} \quad \mathrm{and} \quad \sigma \left(\mathbb{C} ^{2}\right) = \frac{\epsilon _{1} ^{2} + \epsilon _{2} ^{2}}{3} \ ,
\end{equation}
one can read off the functions $ A $ and $ B $ from the exact
deformed partition function as coefficients in the expansion
\begin{equation}
  \label{eq:log-Z-expansion}
\epsilon _{1} \epsilon _{2} \log Z = -F + \left( \epsilon _{1} + \epsilon _{2} \right) H + \epsilon _{1} \epsilon _{2} \log A + \frac{\epsilon _{1} ^{2} + \epsilon _{2} ^{2}}{3} \log B + \cdots \ ,
\end{equation}
where $ \cdots $ are terms that are of higher order in the
$ \Omega $-deformation parameters.

To summarise, our strategy will be very similar to
\cite{Manschot:2019pog}: there are two independent ways of computing
the functions $ A $ and $ B $ --- the first is using the predictions
in \cref{eq:A-and-B-general-forms}, while the second is using
\cref{eq:log-Z-expansion} --- and in comparing the two we will
determine the coefficients $ \alpha $ and $ \beta $. In this manner
\cite{Manschot:2019pog} not only showed that the $ A $ and $ B $
functions are determined by \cref{eq:A-and-B-general-forms}, but also
computed the coefficients $ \alpha $ and $ \beta $ for a number of
rank-$ 1 $ gauge theories. Our focus will be on higher-rank gauge
theories, in particular the so-called $ \mathcal{N} = 2 ^{\star } $
theories (or mass-deformed $ \mathcal{N} = 4 $ super-Yang-Mills
theories) with gauge group $ \mathrm{U}(N) $.

A novel aspect of our analysis will be our use of earlier work on the
resummation of chiral correlators in this class of theories into
quasimodular forms and, in particular, results stemming from the use
of modular anomaly equations in \cite{Ashok:2016ewb}. While we focus
(in the interest of being explicit and concrete) on the case of
$ N = 3 $, it is clear that similar methods may in principle be
employed for any $ N $. We also carry out a perturbative analysis to
determine the scaling dependence of the constants $ \alpha $ and
$ \beta $ on the mass of the adjoint hypermultiplet and the rank of
the gauge group. We also carefully clarify the role of
$ \mathrm{U}(1) $ factors in ensuring that there is consistency and
perfect agreement between our results and those of
\cite{Manschot:2019pog}.

\subsection*{Organisation}
This paper is organised as follows. In \Cref{sec:deform-part-funct} we
study the deformed partition function of $ \mathcal{N} = 2 ^{\star } $
gauge theories with $ \mathrm{U}(N) $ gauge group at a generic point
on the Coulomb branch, first in general, and then specialising to the
case of $ N = 3 $. Here, we read off the $ A $ and $ B $ functions
from the flat-space expansion of the deformed partition function,
writing them in an expansion in the mass $ m $ of the adjoint
hypermultiplet, with coefficients resummed into products of
quasimodular forms of the S-duality group of the theory and functions
of the classical vacuum expectation values of the adjoint scalar
reconstituted into root- and weight-lattice sums. In
\Cref{sec:seib-witt-curv} we compute the $ A $ and $ B $ functions
from predictions using the Seiberg-Witten curve. We then compare the
two to determine the multiplicative coefficients $ \alpha $ and
$ \beta $. In \Cref{sec:u1-factors} we clarify the role of
$ \mathrm{U}(1) $ factors, which are crucial in ensuring perfect
agreement between predictions and microscopic computations. Finally,
in \Cref{sec:pert-analys-mathc} we perform a perturbative analysis of
$ \mathcal{N} = 2 ^{\star } $ theories for all $ N $, taking care to
show that our results are consistent with earlier findings. We
conclude in \Cref{sec:discussion} with a brief summary and some future
directions of study. We collect a few technical details in two
appendices.

\section{Deformed Partition Functions and Localisation}
\label{sec:deform-part-funct}
In this section we discuss the $ \Omega $-deformed partition function $Z$
of the $\mathcal N=2^\star $ gauge theory with gauge group
$ \mathrm{U}(N) $. We will be studying this theory on the Coulomb
branch, where the scalar field $ \phi  $ in the vector multiplet acquires a
vacuum expectation value (vev) that we will parametrise as
\begin{align}
  \label{eq:coulomb-vev-parametrisation}
a=\langle\phi\rangle=\operatorname{diag} \left(a_1,\, \cdots, a_N \right) \ .
\end{align}
The Yang-Mills coupling constant $ g ^{2} $ and the
$ \theta $-angle are packaged, as usual, into the complexified gauge
coupling
\begin{equation}
\tau = \frac{\theta }{2 \pi } + i \frac{4 \pi }{g ^{2}} \ ,
\end{equation}
and, with a view towards its appearance in the instanton sector, it
will be useful to define the elliptic nome
\begin{equation}
q = e^{2 \pi i \tau } \ .
\end{equation}

The partition function of our theory admits a neat factorisation
\cite{Nekrasov:2002qd} into the product of three contributions:
classical (c), perturbative (p) and instanton (np) contributions:
\begin{align}
Z&=Z_{\mathrm{c}} \times Z_{\mathrm{p}} \times Z_{\mathrm{np}}
\end{align}
\subsection{Classical}
The classical contribution is given by 
\begin{align}
\label{Zclass}
 Z_{\text{c}}=q^{-\frac{1}{2\epsilon_1\epsilon_2}\sum_{i}\,a_i^2} \ ,
\end{align}
and this form follows from the classical prepotential that is
quadratic in the $ \mathcal{N} = 2 $ superfield. The appearance of the
$ \left( \epsilon _{1} \epsilon _{2} \right)^{-1} $ is to ensure that
when we do extract the prepotential $ F $ from the deformed partition
function as
\begin{equation}
  \label{eq:prepotential}
F = -\lim_{\epsilon _{1}, \epsilon _{2} \rightarrow 0} \epsilon _{1} \epsilon _{2} \log Z \ ,
\end{equation}
the classical piece contributes as it ought to. Since the $ A $ and
$ B $ functions appear at higher orders in the expansion
\cref{eq:log-Z-expansion} of the deformed partition function, it is
clear at this stage that $ Z _{\mathrm{c}} $ does not contribute to
the functions we are interested in. Consequently, it will not feature
very much in subsequent discussions except for \Cref{sec:u1-factors},
where we will modify \cref{Zclass} in a simple but important way. 

\subsection{Perturbative}
\label{sec:perturbative}
The perturbative contribution is exact at one-loop
\cite{Seiberg:1988ur,Seiberg:1993vc}. In terms of the shorthand
$ \epsilon = \epsilon _{1} + \epsilon _{2} $, the contributions from
the vector multiplet (v) and the adjoint hypermultiplet (h) are given
by \cite{Nekrasov:2002qd,Nekrasov:2003rj}
\begin{equation}
  \label{Zpert}
  \begin{aligned}
Z_{\text{p,v}}&=\prod_{i<j}\text{exp}\left[-\gamma_{\epsilon_1,\epsilon_2}(a_i-a_j;\Lambda)-\gamma_{\epsilon_1,\epsilon_2}(a_i-a_j-\epsilon ;\Lambda)\right] \ , \\
Z_{\text{p,h}}&=\prod_{i,j}\text{exp}\left[\gamma_{\epsilon_1,\epsilon_2}\left(a_i-a_j+m-\epsilon / 2;\Lambda\right)\right] \ .
  \end{aligned}
\end{equation}
Here $\Lambda$ is a cut-off scale and the function $ \gamma _{\epsilon
  _{1},\epsilon _{2}}(x;\Lambda ) $ is defined as
\begin{equation}
\gamma _{\epsilon _{1}, \epsilon _{2}} (x;\Lambda ) = \left. \frac{\mathrm{d} }{\mathrm{d} s} \right\vert _{s=0} \frac{\Lambda ^{s}}{\Gamma (s)} \int _{0} ^{\infty } \mathrm{d} t \, t ^{s-1} \frac{e^{- x t} }{(e^{\epsilon _{1} t} -1)(e^{\epsilon _{2} t} -1)} \ .
\end{equation}

It will be useful to, at this point, take note of a subtlety regarding
the difference between $ \mathrm{U}(N) $ and $ \mathrm{SU}(N) $
theories. In the former, the tensor product of fundamental and
anti-fundamental representations is simply the adjoint, whereas in the
latter we get the trivial representation in addition to the
adjoint. Consequently, in addition to imposing the tracelessness
condition on $ a $ in \cref{eq:coulomb-vev-parametrisation} we must
account for this crucial difference. It may be verified that the
perturbative contributions from the adjoint hypermultiplet in theories
with these two gauge groups are related as
\begin{equation}
Z _{\mathrm{p,h}} \left[ \mathrm{U}(N) \right] = \operatorname{exp} \left[ \gamma _{\epsilon _{1}, \epsilon _{2}}\left( m-\epsilon /2 \right) \right] Z _{\mathrm{p,h}} \left[ \mathrm{SU}(N) \right] \ .
\end{equation}
This will be crucial in ensuring that our results for the dependence
of the constants $ \alpha $ and $ \beta $ on $ N $ and $ m $ are
consistent with those previously derived in the literature, a point we
will return to in \Cref{sec:pert-analys-mathc}. 

\subsection{Nonperturbative}

Let us now discuss the instanton contribution to the partition
function. The instanton partition function is given by the following
contour integral
\cite{Nekrasov:2002qd,Bruzzo:2002xf,Fucito:2011pn,Billo:2012st,Billo:2014bja}
\begin{equation}
Z _{\mathrm{np}} = 1 + \sum_{k=1}^{\infty } \frac{q ^{k}}{k!} \oint \prod _{I=1} ^{k} \frac{\mathrm{d} \chi _{I}}{2 \pi i} \, z _{k} \ ,
\end{equation}
where the $ k $-instanton integrand $ z _{k} $ receives contributions
from the gauge and matter sectors depending on the theory under
consideration. In our case, we have a vector multiplet and an adjoint
hypermultiplet, so
\begin{equation}
z _{k} = z _{k,\mathrm{v}} \times  z _{k,\mathrm{h}}  \ ,
\end{equation}
where
\begin{equation}
  \label{eq:Zkinst}
\begin{aligned}
  z _{k,\mathrm{v}} &= (-1)^{k} \prod _{I,J} ^{k} \left[ \frac{\left(\chi_{IJ}+\delta_{IJ}\right)\left(\chi_{IJ}+\epsilon\right)}{\left(\chi_{IJ}+\epsilon_1\right)\left(\chi_{IJ}+\epsilon_2\right)} \right] \prod _{I = 1} ^{k} \prod _{j=1} ^{N} \left[ -(\chi_I-a_j)^2+ \frac{\epsilon^2}{4} \right] ^{-1} \ , \\
  z _{k,\mathrm{h}} &= \prod _{I,J} ^{k} \left[ \frac{(\chi_{IJ}+\epsilon_1+m)(\chi_{IJ}+\epsilon_2+m)}{(\chi_{IJ}+m)(\chi_{IJ}+\epsilon+m)} \right] \prod_{I=1}^k \prod_{j=1}^N \left[ -(\chi_I-a_j)^2+\frac{(\epsilon+2m)^2}{4} \right] \ .
\end{aligned}
\end{equation}
In the above equations, $\chi_I$ denote the instanton moduli and
$\chi_{IJ}\equiv \chi_I-\chi_J$. The contour integral is computed by
closing the contours in the upper-half $\chi_I$ planes and the
residues are picked using the following pole prescription
\cite{Billo:2012st,Moore:1998et}
\begin{equation}
1 \gg \operatorname{Im} \epsilon_1 \gg \operatorname{Im} \epsilon_2  \gg  0 \ .
\end{equation}

It was emphasised by \cite{Manschot:2019pog} that the $ A $ and $ B $
functions are defined in the Donaldson-Witten twist, and so we must
work with the shifted masses $m\rightarrow m- \epsilon / 2$. This
shift in the mass parameter retains its significance for us as
well. Note that we have already implemented this in the formulas for
the perturbative contributions to the deformed partition function in
\cref{Zpert}. For the instanton sector, all the results that we
report henceforth are obtained \emph{after} performing such a mass
shift in \cref{eq:Zkinst}.

Finally, it is important to keep in mind the global $ \mathrm{U}(1) $
factors that distinguish the partition functions of $ \mathrm{U}(N) $
and $ \mathrm{SU}(N) $ theories. These differences were first
identified in \cite{Alday:2009aq} in the case of rank-$ 1 $ theories
and for $ \mathrm{U}(N) $ theories
\cite{Nekrasov:2015wsu,Jeong:2017mfh,Nekrasov:2017gzb} take the following form:
\begin{equation}
Z _{\mathrm{np}}\left[ \mathrm{U}(N) \right] = Z _{\mathrm{U}(1)} \, Z _{\mathrm{np}} \left[ \mathrm{SU}(N) \right] \ ,
\end{equation}
where
\begin{equation}
  \label{eq:u1-factor-general-N}
Z _{\mathrm{U}(1)} = \left[ \, \prod _{k=1}^{\infty } \left( 1-q ^{k} \right) \, \right]^{-\frac{N}{\epsilon _{1} \epsilon _{2}} (m ^{2} + \epsilon _{1} \epsilon _{2} - \epsilon  ^{2}/4)} \ .
\end{equation}
We mention this in the interest of completeness, should the reader
wish to compare our own results with their own, and to point out that
the Fourier expansion in \cref{eq:u1-factor-general-N} will contribute
nontrivially to the $ A $ and $ B $ functions. It should also be noted
that the dynamics of the low-energy effective theory theory is
insensitive to this difference, since \cref{eq:u1-factor-general-N} is
independent of the Coulomb vevs. Having said that, this contribution
will appear in the microscopic computation of the $ A $ and $ B $
functions, and we will have more to say about it in the following
sections, both in the context of specific examples in \Cref{u3loc},
and in general for arbitrary $ N $ in \Cref{sec:u1-factors}.

We now move onto a discussion of the $ A $ and $ B $ functions
computed via localisation. In the interest of being concrete, we will
present results for the case of $ \mathrm{U}(3) $ gauge
theory. Nevertheless, should one wish to, it is in principle
straightforward to extend these computations to higher $ N $. Further,
the checks coming from curve computations treat all $ \mathrm{U}(N) $
theories on equal footing, so once again explicit checks for arbitrary
$ N $ are in principle possible using the methodology we follow.

\subsection{Localisation Results: $ N = 3 $}
\label{u3loc}

In this section we focus on the gauge group $ \mathrm{U}(3) $ and
compute $ A $ and $ B $ via localisation.

The prepotential $F$ which describes the low-energy effective theory
on the Coulomb branch can in principle be computed to any order in the
instanton expansion from the partition function using
\cref{eq:prepotential}. The resummation of the prepotential into
quasimodular forms of the S-duality group has been extensively
studied; in particular, modular anomaly equations that recursively
determine the prepotential of $ \mathcal{N} = 2 ^{\star } $ gauge
theories order-by-order in a mass expansion have been worked out for
\emph{all} simply laced gauge algebras in \cite{Billo:2015pjb} and
non-simply laced gauge algebras in \cite{Billo:2015jyt}.

At the next order in the expansion \cref{eq:log-Z-expansion} we get
the $H$ function and this turns out to be zero as in the U(2) theory
\cite{Manschot:2019pog}
\begin{align}
H=0 \ .
\end{align}

As we have noted earlier, the classical contribution to the deformed
partition function $ Z _{\mathrm{c}} $ as defined in \cref{Zclass}
does not contribute to either $ A $ or $ B $. Consequently, we will
only focus on the quantum (i.e.~perturbative and instanton)
contributions to both.

\subsubsection{The $ A $ Function}
Let us now consider the contribution to $\log A$ from the perturbative
(p) sector. We make use of the perturbative contribution to the
partition function from the vector multiplet as well as the
hypermultiplet in the adjoint representation as given in
\cref{Zpert}. This yields the following $1$-loop contribution to
$\log A$:
\begin{align}
\log A_{\text{p}}= \frac{1}{2} \sum_{i<j}^{} \log \frac{a _{i} - a _{j}}{\Lambda }  \ .
\end{align}
As for the instanton (np) contributions to $\log A$, we organise them
in a mass expansion:
\begin{align}
\log A_{\text{np}}=\sum_{n=0}A_{n}(a_i,\tau ) \, m^{2n} \ .
\end{align}
The instanton contributions can in fact be resummed into quasimodular
forms of the S-duality group, which in the present case is
$ \mathrm{SL}(2,\mathbb{Z}) $. These forms multiply
functions of the Coulomb vevs, which are most compactly presented in
terms of the following root- and weight-lattice sums, defined in
\cite{Billo:2015pjb,Billo:2015jyt,Ashok:2016ewb} as
\begin{align}
\label{latticesumdef}
C^p_{n;m_1\ldots m_\ell }=\sum_{\lambda\in\mathcal W}\,\sum_{\alpha\in\Psi_{\lambda}}\,\sum_{\beta_1\ne\cdots\ne\beta_\ell \in\Psi_{\alpha}}\frac{(\lambda\cdot a)^p}{(\alpha\cdot a)^n(\beta_1\cdot a)^{m_1}\cdots(\beta_{\ell }\cdot a)^{m_\ell }} \ .
\end{align}
Here, $\mathcal W$ is the set of weights of the fundamental
representation of U($N$), and $\Psi_{\lambda}$ and $\Psi_{\alpha}$ are
the subsets of the root system $\Psi$ defined, respectively, as
\begin{align}
\Psi_{\lambda}=\{\alpha\in\Psi|\lambda\cdot\alpha=1\}\quad\text{for any}\quad\lambda\in\mathcal W \ ,
\end{align}
and
\begin{align}
\Psi_{\alpha}=\{\beta\in\Psi|\alpha\cdot\beta=1\}\quad\text{for any}\quad\alpha\in\Psi \ .
\end{align}
In the interest of keeping the notation as light as possible, we will
also adopt the convention that when $ p = 0 $ in \cref{latticesumdef}
we do not explicitly write the same,
i.e.~$ C ^{0} _{n;m _{1} \cdots m _{\ell }} \equiv C _{n;m _{1} \cdots
  m _{\ell }} $. We refer the reader to Appendix \ref{latticesums} for
more details. 

With these definitions and conventions in place, the results of
equivariant localisation can be resummed order-by-order in the mass of
the adjoint hypermultiplet. Below, we show the first few orders in the
mass expansion of the nonperturbative contribution to the $ A $
function:
\begin{equation}
\label{Ainst}
  \begin{aligned}
\log A_{\text{np}}
&= \left( \frac{3}{2} \, q + \frac{9}{4} \, q ^{2} + 2 \, q ^{3} + \frac{21}{8} \, q ^{4} + O\left(q ^{5}\right)  \right) \\
    &\quad +\frac{m^4}{1152}(8C_4-{C_{2;11}})(E_2^2-E_4)+\frac{m^6}{1080} \bigg[\left(5E_2^3-3E_2E_4-2E_6\right)C_6\ \cr
&\hspace{2.5cm}-\frac{1}{32}\left(5E_2^3-27E_2E_4+22E_6\right)\left(2C_{4;2}+C_{3;3}\right)\bigg]+O(m^8) \ .
  \end{aligned}
\end{equation}
We have not displayed the full Fourier expansions since they are
unwieldy and not particularly enlightening. However, the above
resummed expression in the second and third lines of \cref{Ainst} has
been verified to $ k=4 $ instantons.

What about the first line above, written out explicitly as a Fourier
expansion? The origin of this term is precisely the global
$ \mathrm{U}(1) $ factor we encountered in
\cref{eq:u1-factor-general-N} with $ N=3 $, as can be easily verified
by computing the contribution of $ Z _{\mathrm{U}(1)} $ to the $ A $
function. We will have more to say about this in
\Cref{sec:u1-factors}. For now, we note that this factor won't persist
if we were studying the $ \mathrm{SU}(3) $ theory instead.

\subsubsection{The $ B $ Function}

Let us now consider the contribution to $\log B$ from the perturbative
sector. We make use of the perturbative contribution to the partition
function as given in \cref{Zpert} and obtain the following:
\begin{align}
  \label{eq:log-B-pert}
\log B_{\text{p}}&=\frac{1}{2} \sum_{i < j}^{} \log \frac{a _{i} - a _{j}}{\Lambda } + \frac{1}{8} \sum_{i,j}^{} \log \frac{m + a _{i} - a _{j}}{\Lambda } \ .
\end{align}
For the instanton sector, let us consider the following mass expansion
for $\log B$
\begin{align}
\log B_{\text{np}}=\sum_{n=0}B_{n}(a_i, \tau ) \, m^{2n} \ ,
\end{align}
and, as with the $ A $ function, we find that it is possible to
reconstitute the $ B _{n} (a _{i}, \tau ) $ into quasimodular forms of
the S-duality group multiplying combinations of lattice sums. Taking
both perturbative and nonperturbative terms into account, we find
\begin{equation}
  \label{logB_inst}
  \begin{aligned}
    \log B _{\mathrm{p}+\mathrm{np}}&= -\left( \frac{9}{4} \, q + \frac{27}{8} \, q ^{2} + 3 \, q ^{3} + \frac{63}{16} \, q ^{4} + O \left( q ^{5} \right) \right) \\
    &\quad + \frac{3}{8} \log \frac{m}{\Lambda } + \frac{3}{4} \sum_{i<j}^{} \log \frac{a _{i} - a _{j}}{\Lambda } +\frac{m^2}{64}C_{2;11}\left(3C^2-\left(C^1\right)^{2}\right)E_2 \\
&\quad  -\frac{m^4}{256}\Big[4\left(E_2^2+E_4\right)C_4+\left(E_2^2-E_4\right)C_{2;11}\Big] \\
&\quad -\frac{m ^{6}}{34560} \Big[8\left(25E_2^3+48E_2E_4+17E_6\right)C_6 +24\left(5E_2^3+3E_2E_4-8E_6\right)C_{5;1} \\
&\hspace{3cm}-5\left(5E_2^3-3E_2E_4-2E_6\right)C_{2;22}\Big]+O(m^8) \ .
  \end{aligned}
\end{equation}
Once again, we have elected not to explicitly write out the Fourier
series.\footnote{Note that \cref{logB_inst} contains both perturbative
  and nonperturbative contributions. Therefore, at weak coupling
  (i.e.~$ q \rightarrow 0 $) it reduces to the small $ m $ expansion
  of \cref{eq:log-B-pert}.}  Nevertheless, as in the case of $\log A$,
we have verified the above resummed expression up to $ k=4 $
instantons. Further, the first line in the above equation comes from
the $ \mathrm{U}(1) $ factor, as can be verified by computing the
contribution of $ Z _{\mathrm{U}(1)} $ in
\cref{eq:u1-factor-general-N} to $ \log B $. We will have more to say
about this factor in \Cref{sec:u1-factors}. For now, we note that had
we elected to study the $ \mathrm{SU}(3) $ theory instead, this factor
would not have been present.

\section{Seiberg-Witten Curves and Testing Predictions}
\label{sec:seib-witt-curv}
We'd like to compute the $ A $ and $ B $ functions using
\cref{eq:A-and-B-general-forms}. For this, we'll need to compute the
chiral correlators $ u _{k} = \left\langle \operatorname{Tr} \Phi ^{k}
\right\rangle $ and the physical discriminant $ \Delta  $. We compute
these quantities using the corresponding Seiberg-Witten curve. On
comparing these results with those of the previous section, we should
be able to fix the constants $ \alpha  $ and $ \beta  $. 

Before specialising to the case of $ N = 3 $, we briefly discuss (for
general $ N $) what the Seiberg-Witten curve of these theories looks
like. For our purposes, the Donagi-Witten \cite{Donagi:1995cf} form of
the curve is most suitable, but it will be useful to understand
different parametrisations and how they are related. In the
Donagi-Witten formulation, the Seiberg-Witten curve of an
$ \mathrm{U}(N) $ gauge theory with a massive adjoint hypermultiplet
is given by an $N$-fold cover of an elliptic curve of the form
\begin{align}
\label{g1curve}
y^2=(x-\mathfrak{e}_1)(x-\mathfrak{e}_2)(x-\mathfrak{e}_3) \ ,
\end{align}
where the $\mathfrak{e}_i$ sum to zero and their pairwise differences are
proportional to Jacobi $ \theta $-constants (see Appendix
\ref{Modforms} for more details):
\begin{equation}
\begin{aligned}
  \mathfrak{e} _{2} - \mathfrak{e} _{3} &= \frac{1}{4} \theta _{2} (\tau )^{4} \ , \\
  \mathfrak{e} _{2} - \mathfrak{e} _{1} &= \frac{1}{4} \theta _{3} (\tau )^{4} \ , \\
  \mathfrak{e} _{3} - \mathfrak{e} _{1} &= \frac{1}{4} \theta _{4} (\tau )^{4} \ .
\end{aligned}
\end{equation}
Expanding out \cref{g1curve} and making use of the relation between
the Jacobi $\theta$-constants and the Eisenstein series (reviewed in
\Cref{Modforms}), one finds that the elliptic curve can also be
written as
\begin{equation}
y ^{2} = x ^{3} - \frac{E _{4}}{48} x + \frac{E _{6}}{864} \ .
\end{equation}
Now, under modular transformations we know that $ E _{2k} $ has weight
$ 2k $. For consistency, we must assign modular weights to $ x \ (2) $ and
$ y \ (3) $ as well. 

The $ N $-fold cover of this curve takes the form
\begin{equation}
  \label{eq:sw-curve-general-N}
F(t,x,y) = \sum_{n=0}^{N} (-1)^{n} A _{n} \, P _{N-n}(t,x,y) = 0 \ .
\end{equation}
For consistency, $ t $ is assigned unit modular weight. The quantities
$ A _{n} $ are chiral correlators that transform in a simple manner
(covariantly, as weight-$ n $ objects) under S-duality transformations
and parametrise the Coulomb moduli space, and the $ P _{n} $ are
polynomials that are determined recursively using
\begin{equation}
\frac{\mathrm{d} P _{n}}{\mathrm{d} t} = n P _{n-1} \ ,
\end{equation}
together with conditions on the growth of these polynomials near
infinity. We refer the reader to \cite{Ashok:2016ewb} for a more
detailed discussion of the curve and its different forms.

Before proceeding, we must clarify the manner in which the
coefficients $ A _{k} $ that appear in the Seiberg-Witten curve are
related to the gauge invariant Coulomb moduli $ u _{k} $, since we
have closed-form expressions for the former (in a mass expansion)
whereas the latter are related to the $ A $ function via
\cref{eq:A-and-B-general-forms}.

\subsection{Parametrising the Coulomb Branch}
Different forms of the Seiberg-Witten curve entail different
parametrisations of the Coulomb moduli space. For example, we might
choose to work with the D'Hoker-Phong \cite{DHoker:1997hut} curve:
\begin{equation}
H(t) = \prod _{i = 1} ^{N} (t - e _{i}) = \sum_{n=0}^{N} (-1)^{n} W _{n} t ^{N-n} \ ,
\end{equation}
where the $ e _{i} $ are understood to be quantum corrected Coulomb
vevs in that at very weak coupling they reduce to $ a _{i} $. In terms
of these, the $ W _{n} $ are
\begin{equation}
  \label{eq:dhoker-phong-wn}
W _{n} = \sum_{i _{1} < \cdots < i _{n}}^{} e _{i _{1}} \cdots e _{i _{n}} \ ,
\end{equation}
and it is easy to see that such quantities can be used to build up our
object of interest (the chiral correlators $ u _{k} $) since
\begin{equation}
  \label{eq:coulomb-moduli-un}
  u _{k} =  \left\langle \frac{1}{k} \operatorname{Tr} \Phi ^{k} \right\rangle = \frac{1}{k} \sum_{i=1}^{N}  e _{i} ^{k} \ .
\end{equation}
Our task in this section will be to review the manner in which these
three different parametrisations of the Coulomb branch are related to
each other.

First, the D'Hoker-Phong $ (W _{n}) $ and Donagi-Witten $ (A _{n}) $
parametrisations are related linearly, and the map between the two is
determined solely by quasimodular forms \cite{Ashok:2016ewb}. The
relation between the two is
\begin{equation}
  \label{eq:a-w-expansion}
A _{n} = \sum_{\ell = 0}^{\lfloor n/2 \rfloor} \binom{N-n+2 \ell }{2 \ell } \left( 2 \ell -1 \right)!! \left( \frac{m ^{2}E _{2}}{12}  \right)^{\ell } W _{n-2 \ell } \ ,
\end{equation}
which admits a simple inversion:
\begin{equation}
  \label{eq:w-a-expansion}
W _{n} = \sum_{\ell = 0}^{\lfloor n/2 \rfloor } \left( -1 \right)^{\ell } \binom{N-n+2 \ell }{2 \ell } \left( 2 \ell -1 \right)!! \left( \frac{m ^{2}E _{2}}{12}  \right)^{\ell } A _{n-2 \ell } \ .
\end{equation}

Next, let us review the relation between elementary symmetric
polynomials constructed out of the quantum corrected Coulomb vevs in
\cref{eq:dhoker-phong-wn} and their corresponding power sums in
\cref{eq:coulomb-moduli-un}. The relation between the two can be
written as a logarithmic generating function:
\begin{equation}
  \label{eq:uk-vs-wk-generating-function}
\sum_{k=1}^{\infty } (-1)^{k-1} \, t ^{k} \, u _{k} = \log \left[ 1 + \sum_{k=1}^{\infty } t ^{k} \, W _{k} \right] \ ,
\end{equation}
which, on comparing terms, gives
\begin{equation}
\begin{aligned}
  u _{1} &= W _{1} \ , \\
  u _{2} &= \frac{1}{2} \left( W _{1}^{2} - 2 W _{2} \right) \ , \\
  u _{3} &= \frac{1}{3} \left( W _{1}^{3} - 3 W _{1} W _{2} + 3 W _{3} \right) \ ,
\end{aligned}
\end{equation}
and so on. These relations will be crucial for us, since the
prediction for the $ A $ function in \cref{eq:A-and-B-general-forms}
is given in terms of the $ u _{k} $ while earlier work on resummation
and S-duality naturally finds closed-form expressions (in a mass
expansion) for the $ W _{k} $ or $ A _{k} $. To proceed, we will use
\cref{eq:uk-vs-wk-generating-function} to arrive at closed-form
expression for the $ u _{k} $ before computing $ A $ via
\cref{eq:A-and-B-general-forms}.

For the rest of this section, in order to compare with the results of
\Cref{u3loc}, we specialise to the case of $ N = 3 $.

\subsection{Chiral Correlators and the $ A $ Function}
Setting $ N = 3 $ in \cref{eq:sw-curve-general-N} and using the
expressions for the $ P _{n} $ from \cite{Ashok:2016ewb}, the curve
for the $ \mathcal{N} = 2 ^{\star } $ theory with gauge group
$ \mathrm{U}(3) $ can be rearranged as
\begin{equation}
\label{curve1}
F(t,x,y) = t^3 - A_1 t^2 + \left(A_2 -3m^2 x\right) t -\left(A_3-m^2 A_1 x- 2 m^3 y\right) = 0 \ .
\end{equation} 
where $A_k$ are gauge invariant coordinates on the Coulomb moduli
space of the theory. The three independent $A_k$ for the
$ \mathrm{U}(3) $ theory \cite{Ashok:2016ewb} are given in terms of
the root- and weight-lattice sums in \cref{latticesumdef} as
\begin{equation}
  \begin{aligned}
\label{Aicoord}
A_1 &= C ^{1} \ , \\
A_2 &= \frac{1}{2!} \left( \left( C ^{1} \right)^{2} - C ^{2} \right) 
+\frac{m^2}{4}\,E_2+ \frac{m^4}{288}\,\big(E_2^2-E_4\big)\,C_{2}
+\frac{m^6}{4320}\,\big(5E_2^3-3E_2E_4-2E_6\big)\,C_{4}\\
&\quad\quad+\frac{m^6}{3456}\,\big(E_2^3-3E_2E_4+2E_6\big)\,C_{2;11} + O(m^8) \ , \\
A_3 &= \frac{1}{3!} \left( \left( C ^{1} \right)^{3} - 3 C ^{2}C ^{1} + 2 C ^{3} \right)
+\frac{m^2}{12}\, E_2 \, C ^{1} +\frac{m^4}{288}\,\big(E_2^2-E_4\big)
\Big(C_{2}\, C ^{1}
- 2\, C^{1}_{2} \Big)\\
&\quad\quad +\frac{m^6}{4320}\,\big(5E_2^3-3E_2E_4-2E_6\big)\Big(C_{4} 
\,C ^{1}-2\, C^{1}_{4}\Big)\\
&\quad\quad +\frac{m^6}{3456}\,\big(E_2^3-3E_2E_4+2E_6\big)\Big(C_{2;11} 
\,C ^{1}-2\, C^{1}_{2;11}\Big)+O(m^8) \ , 
  \end{aligned}
\end{equation}

As we reviewed in the Introduction, the $A$ function is expected to be
\cite{Moore:1997pc,Losev:1997tp,Marino:1998bm,Shapere:2008zf,Witten:1995gf}
\begin{align}
\label{relationAB}
A = \alpha \left(\operatorname{det}\,\frac{\mathrm{d} u_i}{\mathrm{d} a_j}\right)^{1/2} \ .
\end{align}
We can now use \cref{Aicoord} together with
\cref{eq:w-a-expansion,eq:uk-vs-wk-generating-function} to compute the
$ u _{k} $ and subsequently the $A$ function according to
\cref{relationAB}. We computed the logarithm of this function and
checked that the resulting expression matches the results for the same
obtained via localisation in \Cref{u3loc} provided
\begin{align}
\alpha=f _{\alpha }\,\Lambda^{-3/2} \ ,
\end{align}
where $ f _{\alpha } $ is some as-yet-undetermined function of the
complexified gauge coupling $ \tau $ that will ensure that the
nonperturbative contribution from the $ \mathrm{U}(1) $ factor
matches. Said differently, without $ f _{\alpha } $ in the above
equation there is a mismatch between $ \log A $ computed using
\cref{relationAB} and the same function computed microscopically as in
\Cref{u3loc}, and this mismatch is essentially due to the
$ \mathrm{U}(1) $ factor in (\cref{eq:u1-factor-general-N}). We will
fix $ f _{\alpha } $ in \Cref{sec:u1-factors}.

Note that the analysis of \cite{Ashok:2016ewb} treats all
$ \mathrm{U}(N) $ gauge theories on equal footing, and it is in
principle possible to perform a similar computation for any $ N
$. Indeed, in \Cref{sec:pert-analys-mathc} we will perform a
perturbative analysis that will determine the scaling of $ \alpha $
with $ N $.

We now move onto a discussion of the $ B $ function.

\subsection{Discriminants and the $ B $ Function}
Having demonstrated how one might go about computing the $ A $
function for a higher-rank gauge theory, we now turn to the
computation of the discriminant. The strategy we adopt is to eliminate
the variables $x$ and $y$ in \cref{curve1} and write down a
polynomial equation in $t$. After taking care of a few subtleties, the
discriminant of this polynomial is the relevant ``physical''
discriminant.

We eliminate $y$ from \cref{curve1} using \cref{g1curve}. This
substitution leaves us with the following polynomial in $(t,x)$,
\begin{align}
\label{curve2}
Q = \big(t^3 - A_1 t^2 +t (A_2 -3m^2 x) -A_3 + m^2 A_1 x\big)^2 - 4 m ^{6} (x-\mathfrak{e}_1)(x-\mathfrak{e}_2)(x-\mathfrak{e}_3) \ .
\end{align}
The discriminant of the Seiberg-Witten curve captures, as usual, the
singular locus of the curve. Singularities arise when the following
conditions are satisfied
\begin{align}
 \frac{\partial Q}{\partial t}= \frac{\partial Q}{\partial x}=Q=0 \ .
\end{align}
The first of these requirements leads to
\begin{align}
x = \frac{3 t^2 - 2 t A_1 +A_2}{3 m^2}
\end{align}
and we use this to eliminate $x$ from $Q $ in
\cref{curve2}.\footnote{This could also imply
  $t^3 - 3 m^2 t x - t^2 A_1 + m^2 x A_1 + t A_2 - A_3 = 0$. However,
  this condition is not compatible with $Q = \partial _{x} Q =0$
  when all the $\mathfrak{e}_i$ are distinct.} This yields a
polynomial $H(t)$. 
The singularities of the curve should satisfy the conditions
$H = \partial _{t}H =0$. This, however, does not imply that all the
zeroes of these equations correspond to singularities of the
curve. For example, in this case we see that $t = A _{1} / 3$, which
is not a singularity of the curve, is a solution of
$\partial _{t} H=0$. Thus, when we compute the discriminant of the
polynomial $H(t)$, we discard this factor and refer to the remaining
part as the ``physical'' discriminant $\Delta$ that is sensitive only
to the singularities associated to massless particles. The factor that
ought to be discarded is an overall multiplicative factor in the
``mathematical'' discriminant and hence can be removed
easily. Further, as in \cite{Donagi:1995cf}, the explicit form of
$\Delta$ is quite complicated but in the weak coupling limit and with
tracelessness condition $\sum_{i}\,a_i = 0 $ it reduces to
\begin{align}
(4 A_2^3 +27 A_3^2)^2 ~\left(4 (A_2-m^2)~(A_2-4m^2)^2+27 A_3^2\right) \ .
\end{align}

The $B$ function is expected to be related to the discriminant as
\cite{Moore:1997pc,Losev:1997tp,Marino:1998bm,Shapere:2008zf,Witten:1995gf}
\begin{align}
\label{relationAB2}
B= \beta \Delta^{1/8} \ .
\end{align}
Once we have the correct discriminant $\Delta$, we can easily compute
the $B$ function according to \cref{relationAB2}. We computed the
logarithm of this function and checked that we have agreement between
the two computations provided we set
\begin{align}
\beta=f _{\beta } \, m^{\frac{3}{8}}\,\Lambda^{-\frac{21}{8}} \ ,
\end{align}
where, just as in the case of the $ A $ function, the factor
$ f _{\beta } $ in the above equation is some as-yet-undetermined
function of the complexified gauge coupling $ \tau $ that will ensure
that the nonperturbative contribution from the $ \mathrm{U}(1) $
factor matches. We will fix $ f _{\beta } $ in the next section. 

In principle, as with the previous section, the procedure followed in
this section can be repeated for higher-rank gauge theories.

\section{$ \mathrm{U}(1) $ Factors}
\label{sec:u1-factors}
In this section we discuss in greater detail the $ \mathrm{U}(1) $
factors and the manner that they affect determination of coefficients
$ \alpha $ and $ \beta $ associated to the $ A $ and $ B $ functions
through the as-yet-unspecified functions $ f _{\alpha } $ and
$ f _{\beta } $ we saw in the previous section. Our discussion of
these factors is influenced by the treatment of $ \mathrm{U}(1) $
factors for gauge theories with rank-$ 1 $ gauge groups in
\cite{Manschot:2021qqe} and our goal in this section is to carry out a
similar program for the higher-rank gauge theories we discuss in this
paper.\footnote{We are grateful to Jan Manschot for emphasising this.}

Let us begin by rewriting the $ \mathrm{U}(1) $ factor in
\cref{eq:u1-factor-general-N} in terms of the Dedekind $ \eta
$-function (see \Cref{Modforms} for its definition) as
\begin{equation}
Z _{\mathrm{U}(1)} = \left[ \frac{q ^{1/24}}{\eta (\tau )}  \right]^{\frac{N}{\epsilon _{1} \epsilon _{2}} \left( m ^{2}+\epsilon _{1}\epsilon _{2} - \epsilon ^{2}/4 \right)} \ .
\end{equation}
We can compensate for the explicit appearance of the elliptic nome $ q
$ in the above expression by starting with a classical partition
function that is slightly different. That is, had we started with
\begin{equation}
  \label{eq:modified-z-class}
Z _{\mathrm{c}'} = Z _{\mathrm{c}} \times q ^{-\frac{N}{24 \epsilon _{1} \epsilon _{2}} \left( m ^{2} + \epsilon _{1} \epsilon _{2} - \epsilon ^{2}/4 \right)} \ ,
\end{equation}
where $ Z _{\mathrm{c}} $ is in \cref{Zclass}, we will be able to
express the effect of the $ \mathrm{U}(1) $ factor on the constants
$ \alpha $ and $ \beta $ solely in terms of the Dedekind
$ \eta $-function, as in \cite{Manschot:2021qqe}. Note that this
redefinition does not affect the dynamics of the theory, since it is
independent of the Coulomb vevs.

Assuming that the classical piece chosen is $ Z _{\mathrm{c}'} $ in
\cref{eq:modified-z-class}, we can now expand the logarithm of the
product of modified classical $ (c') $ and $ \mathrm{U}(1) $ factors
around flat space as we did with \cref{eq:log-Z-expansion}, and from
this we can read off its contribution to the prepotential $ F $, as
well as the $ A $ and $ B $ functions. Focusing solely on the
non-dynamical pieces, i.e.~pieces that are independent of the Coulomb
vevs, we find:
\begin{equation}
  \label{eq:modified-class-u1-together}
\epsilon _{1} \epsilon _{2} \log \left[ Z _{\mathrm{c}'} Z _{\mathrm{U}(1)} \right] \bigg\vert _{a _{i} = 0} = -N m ^{2} \log \eta (\tau ) - N \left( \epsilon _{1} \epsilon _{2} - \frac{\epsilon ^{2}}{4}  \right) \log \eta (\tau ) \ .
\end{equation}
The first term on the right-hand side in the above equation is
precisely the $ a $-independent factor that was neglected in
\cite{Billo:2015pjb}. From the second term we can read off the
contribution of the $ \mathrm{U}(1) $ factor to the $ A $ and $ B $
functions of the $ \mathrm{U}(N) $ theories. We find after some simple
algebra that
\begin{equation}
  \label{eq:f-alpha-beta-general-N}
  \begin{aligned}
f _{\alpha } &= \eta (\tau )^{-\frac{N}{2} } \ , \\
f _{\beta } &= \eta (\tau )^{+\frac{3N}{4} } \ .
  \end{aligned}
\end{equation}
For the case of $ N = 2 $, these match the results of
\cite{Manschot:2021qqe} exactly.\footnote{The results of
  \cite{Manschot:2021qqe} include a third coupling $ C $; this is
  essentially the contribution of the $ \mathrm{U}(1) $ factor to the
  prepotential $ F $ of the gauge theory. For $ N=2 $, the power of
  $ \eta $ corresponding to $ C $ matches too, as can be read off from
  the first term on the right-hand side of
  \cref{eq:modified-class-u1-together}.}

\section{Perturbative Analysis of $\mathcal N=2^\star $ $ \mathrm{U}(N) $ Theories}
\label{sec:pert-analys-mathc}
In this section, we will compute the perturbative contributions to the
$A$ and $B$ functions in $\mathcal N=2^\star $ theories with U($N$)
gauge group where $N$ is arbitrary. We will then use this to determine
the dependence of the constants $\alpha$ and $\beta$ on the rank $ N $
of the gauge group, the ultraviolet cut-off scale $ \Lambda $, and the
mass $ m $ of the adjoint hypermultiplet.\footnote{It was argued in
  \cite{Marino:1998tb} that for asymptotically free theories $\alpha$
  and $\beta$ are independent of mass parameter. This doesn't apply,
  however, to the theory we are presently discussing.} Little is known
about these constants for $\mathcal N=2^\star $ theory except for the
SU($2$) gauge group \cite{Manschot:2019pog,Labastida:1998sk} and we
will use their results as a consistency check.

The perturbative contribution to the partition function is given in
\cref{Zpert}. If we expand \cref{Zpert} around the flat space limit
i.e.~$\epsilon_{1,2} \to 0$, we obtain the following perturbative
contribution
\begin{equation}
  \begin{aligned}
\label{pert_contri}
\log A _{\text{p}} &= \frac{1}{2} \sum_{ i < j } \log \left(\frac{a_i-a_j}{\Lambda}\right) \ , \\
\log B _{\text{p}} &= \frac{1}{2} \sum_{ i < j } \log \left(\frac{a_i-a_j}{\Lambda}\right) +\frac{1}{8} \sum_{i,j} \log \left(\frac{m+a_i-a_j}{\Lambda}\right) \ .
  \end{aligned}
\end{equation}
Let us now compute the perturbative contribution to the $A$ and $B$
functions from the SW curve. The perturbative contribution to the
$u_k$ variables is given by \cref{eq:coulomb-moduli-un} in the limit $
q \rightarrow 0 $, which is the limit in which the quantum corrected
Coulomb vevs match their classical values, so
\begin{align}
u_k \big\vert _{\mathrm{p}}= \frac{1}{k} \sum_{i}^{} a _{i}^{k}  \ .
\end{align}
The entries of the matrix appearing in \cref{relationAB} are then
straightforwardly found to be
\begin{align}
\left( \frac{\mathrm{d} u_i}{\mathrm{d} a_j}\right)_{\text{p}} = a _{j} ^{\ i-1}  \ .
\end{align}
This matrix is well-known --- it is the Vandermonde matrix --- and so
its determinant takes the familiar form
\begin{align}
  \label{eq:perturbative-du-da-matrix}
\det\left( \frac{\mathrm{d} u_i}{\mathrm{d} a_j}\right)_{\text{p}} = \prod_{  i<j } \left(a_i - a_j\right) \ .
\end{align}
Next, the perturbative discriminant receives a contribution due to the
singularities coming from both the vector and adjoint hypermultiplet
\cite{Donagi:1995cf} in the form
\begin{equation}
  \label{eq:perturbative-discriminant}
  \Delta_{\text{p}} = \Delta_{\mathrm{p},\mathrm{v}}^2 \, \Delta_{\mathrm{p},\mathrm{h}} \ ,
\end{equation}
where
\begin{equation}
  \begin{aligned}
\Delta_{\mathrm{p},\mathrm{v}} &= \prod_{i<j} (a_i-a_j)^2 \ , \\
\Delta_{\mathrm{p},\mathrm{h}} &= \prod_{i\ne j} (a_i-a_j+m)= (-1)^{N(N+1)/2}~ \prod_{i<j} \left((a_i-a_j)^2-m^2\right) \ .
  \end{aligned}
\end{equation}

Using \cref{eq:perturbative-du-da-matrix,eq:perturbative-discriminant}
we can compute $ \log A $ and $ \log B $ using
\cref{relationAB,relationAB2}. We then compare this with
\cref{pert_contri}, which allows us to determine the dependence of the
constants $ \alpha  $ and $ \beta  $ on the rank $ N $ of the gauge
group and the mass $ m $ of the adjoint hypermultiplet as
\begin{align}
\alpha _{\mathrm{U}(N)} = \Lambda^{-\frac{N(N-1)}{4}} \quad \mathrm{and} \quad  \beta _{\mathrm{U}(N)} = m^{\frac{N}{8}}~\Lambda^{-\frac{N(3N-2)}{8}} \ .
\end{align}
As a consistency check, we can ask if the above expressions match the
results of \cite{Labastida:1998sk,Manschot:2019pog} for the case of
gauge group $ \mathrm{SU}(2) $. To see this, recall that in the case
of $ \mathrm{SU}(N) $ gauge groups the perturbative contribution to
the $ \Omega $-deformed partition function from the adjoint
hypermultiplet in \cref{Zpert} should be divided by a factor of
$\exp
\left[\gamma_{\epsilon_1,\epsilon_2}(m-\epsilon/2;\Lambda)\right]$, as
was discussed in \Cref{sec:perturbative}. This leaves $\log A$
unchanged but affects $\log B$ and correspondingly we find
\begin{align}
\beta_{\text {SU} (N)} = m^{\frac{N-1}{8}}~\Lambda^{-\frac{(3N+1)(N-1)}{8}} \ .
\end{align}

For the case of $ N = 2 $, the above formula correctly reproduces the
scaling for $\beta$ in the $ \mathcal{N} = 2 ^{\star } $ theory with
$ \mathrm{SU}(2) $ gauge group (upto some numerical factors) as given
in \cite{Manschot:2019pog}.

\section{Conclusions}
\label{sec:discussion}
In this paper, we studied the low-energy effective couplings $A$ and
$B$ of $\mathcal N=2 $ supersymmetric gauge theories to the
topological invariants of a four-manifold, which in our case was the
$ \Omega $-background. As in \cite{Manschot:2019pog}, we used the
deformed partition function on this background to explicitly calculate
the quantities $A$ and $B$. We then compared this with results from
the corresponding Seiberg-Witten curves, taking particular care to
treat $ \mathrm{U}(1) $ factors. Our findings generalise the results
of \cite{Manschot:2019pog} to $ \mathcal{N}=2 ^{\star } $ theories
with higher-rank $ \mathrm{U}(N) $ gauge groups. While we have
explicitly worked out the case of $ N=3 $, it is clear that the
methodology we have employed will continue to work for higher $ N
$. To complement our findings, we have also presented a perturbative
study of the the $A$ and $B$ functions in these theories and
determined the manner in which the multiplicative constants $\alpha$
and $\beta$ scale with the rank of the gauge group and the mass of the
adjoint hypermultiplet.

Looking ahead, it would be interesting to explore the possibility of
using alternative methods --- such as the Eynard-Orantin topological
recursion \cite{Eynard:2007kz} or the theory of qq-characters
\cite{Nekrasov:2015wsu,Nekrasov:2016qym,Nekrasov:2016ydq,Nekrasov:2017rqy,Nekrasov:2017gzb}
--- to derive all-instanton results. We also note that
\cite{Manschot:2019pog} considered the case of the $\mathrm{SU}(2)$
theory with $N_f=4$ fundamental hypermultiplets. The extension of this
analysis to superconformal SQCD theories with gauge group
$\mathrm{SU}(N)$ and $N_f =2N$ fundamental hypermultiplets is more
complicated. Nevertheless, there are results on co-dimension $ 1 $
loci for $ N=3 $ in \cite{Ashok:2015cba,Aspman:2020lmf} and results
for general $ N $ in \cite{Ashok:2016oyh}, which may provide a
suitable starting point for such an analysis. Another possible
direction would be the extension to other gauge algebras, where
accidental isomorphisms at low rank will provide consistency
checks. We hope to return to some of these questions in the future.

\section*{Acknowledgments}
We thank Sujay K. Ashok for useful discussions, especially related to
the discriminant of the Seiberg-Witten curve, and feedback on an
earlier draft. We are also grateful to Jan Manschot for bringing to
our attention \cite{Manschot:2021qqe}, which helped improve our
discussion of the $ \mathrm{U}(1) $ factors. We thank Marco Bill\'o
and Eleonora Dell'Aquila for access to their localisation
computations. RRJ thanks IISER Pune for support and hospitality during the time when part of this work was done. SM thanks IIT Indore for hospitality during the ST$^4$ workshop where part of this
work was done. MR is supported by Grant No. 21/02253-0 and 19/21281-4,
São Paulo Research Foundation (FAPESP).

\appendix
\section{Lattice Sums}
\label{latticesums}
For the unitary gauge group $ \mathrm{U}(N) $ the root system of the
corresponding gauge algebra $ \mathfrak{u}(N) $ is given in terms of
an set of vectors $ \{\texttt{e} _{i}\} $ in $ \mathbb{R}^{N} $ where
$ i $ runs over $ \{1, \cdots , N\} $ such that
$ \texttt{e}_{i} \cdot \texttt{e}_{j} = \delta _{ij} $, i.e.~the basis
is orthonormal. In terms of this, the roots are the set
\begin{equation}
\left\lbrace \pm \left( \texttt{e}_{i} - \texttt{e}_{j} \right) \ \big\vert \ 1 \leq i < j \leq N \right\rbrace \ .
\end{equation}
This basis can be used to expand out the lattice sums in
\cref{latticesumdef}.

For example:
\begin{equation}
C _{n} = \sum_{i \neq j}^{} \frac{1}{\left(a _{i} - a _{j}\right) ^{n}} \ ,
\end{equation}
which, as is easily verified, is identically zero for $ n $ odd. As
another example, for $ N=3 $ we have:
\begin{equation}
C _{2;11} = 4\,\frac{a _{1}a _{2} + a _{1} a _{3} + a _{2} a _{3} - a _{1} ^{2} - a _{2} ^{2} - a _{3} ^{2}}{\left( a _{1} - a _{2} \right)^{2}\left( a _{1} - a _{3} \right)^{2}\left( a _{2} - a _{3} \right)^{2}} \ .
\end{equation}

\section{Modular Forms}
\label{Modforms}
In this appendix, we give some details of the modular forms that
appeared in the main text of the paper.

The Dedekind $ \eta  $-function is defined in terms of an infinite
product as
\begin{equation}
\eta (\tau ) = q ^{1/24} \prod _{k=1} ^{\infty } \left( 1-q ^{k} \right) \ .
\end{equation}

The Jacobi $\theta$-functions
are defined as
\begin{equation}
 \theta\left[^a_b\right](v\vert \tau) = \sum_{n\in\mathbb Z} \mathrm{e}^{\pi i\tau \left(n - \frac{a}{2}\right)^2
+ 2\pi i\left(n - \frac{a}{2}\right)\left(v-\frac{b}{2}\right)}~,
\end{equation}
for $a,b=0,1$. We also have Jacobi $\theta$-constants that are defined
as follows
\begin{align}
\theta_2(\tau)\equiv\theta\left[^1_0\right](0\vert \tau),\quad
\theta_3(\tau)\equiv\theta\left[^0_0\right](0\vert \tau),\quad
\theta_4(\tau)\equiv\theta\left[^0_1\right](0\vert \tau)
\end{align}
The first few terms in the Fourier expansion of these
$\theta$-constants are as follows
\begin{equation}
  \begin{aligned}
\theta_2(\tau)&=2q^{\frac 18}\left(1+q+q^3+q^6+\cdots\right) \ , \\
\theta_3(\tau)&=1+2q^{\frac 12}+2q^2+2q^{\frac 92}+2q^8+\cdots \ , \\
\theta_4(\tau)&=1-2q^{\frac 12}+2q^2-2q^{\frac 92}+2q^8+\cdots \ .
  \end{aligned}
\end{equation}
where $q=e^{2\pi i\tau}$.

The Eisenstein series $ E _{2k} $ are holomorphic functions on the
upper-half plane, defined as
\begin{equation}
E _{2k}(\tau ) = \frac{1}{2 \zeta (2n)} \sum_{m,n \in \mathbb{Z}^{2}\backslash\{0,0\}}^{} \frac{1}{\left( m + n \tau  \right)^{2k}} \ ,
\end{equation}
which makes explicit that under an $ \mathrm{SL}(2,\mathbb{Z}) $
transformation of the argument, the Einstein series transform
covariantly with weight $ 2k $. In terms of the generators $ T $ and
$ S $ of the modular group that act on the upper-half plane coordinate
$ \tau $ via fractional linear transformations, the Eisenstein series
(for $ k \geq 2 $) transform as
\begin{equation}
\begin{aligned}
  T \quad \colon \quad  E _{2k} (\tau +1) &= E _{2k}(\tau ) \ , \\
  S \quad \colon \quad E _{2k} (-1/\tau ) &= \tau ^{2k} \, E _{2k}(\tau ) \ .
\end{aligned}
\end{equation}
In particular, the $ T $-invariance of the Eisenstein series tells us
that they admit a Fourier series; the Fourier expansion of the first
few Eisenstein series are
\begin{equation}
  \begin{aligned}
E_4(\tau)&=1+240q+2160q^2+6720q^3+17520q^4+\cdots\ , \\
E_6(\tau)&=1-504q-16632q^2-122976q^3-532728q^4+\cdots \ .
  \end{aligned}
\end{equation}
The case of $ k=1 $ is special: under an $ S $ transformation, it
transforms anomalously as
\begin{equation}
E _{2}(-1/\tau ) = \tau ^{2} E _{2} (\tau ) + \frac{6}{i \pi } \, \tau \ ,
\end{equation}
and is therefore referred to as a quasimodular form. It, too, has a
Fourier expansion:
\begin{equation}
E _{2} (\tau ) = 1 - 24 q - 72 q ^{2} - 96 q ^{3} - 168 q ^{4} - \cdots \ .
\end{equation}

The Eisenstein series of weights $ 4 $ and $ 6 $ can be written as
polynomials in the Jacobi $ \theta  $-constants:
\begin{equation}
\begin{aligned}
  E _{4}(\tau ) &= \frac{1}{2} \left( \theta _{2}(\tau )^{8}+ \theta _{3}(\tau )^{8} + \theta _{4}(\tau )^{8} \right) \ , \\
  E _{6}(\tau ) &= \frac{1}{2} \left( \theta _{3} (\tau )^{4}+\theta _{4}(\tau )^{4} \right)\left( \theta _{2} (\tau )^{4}+\theta _{3}(\tau )^{4} \right)\left( \theta _{4} (\tau )^{4}-\theta _{2}(\tau )^{4} \right)
\end{aligned}
\end{equation}

\providecommand{\href}[2]{#2}\begingroup\raggedright
\bibliography{refsAB}
\bibliographystyle{JHEP}
\endgroup

\end{document}